\documentclass[10pt,a4paper]{article}
\usepackage{amsmath}
\usepackage{amssymb}
\usepackage{amsfonts}
\usepackage{amstext}
\usepackage{amsbsy}
\usepackage[mathscr]{eucal}
\usepackage{graphicx}
\usepackage{color}
\usepackage[all]{xy}
\newcommand{\beq}{\begin{equation}}
\newcommand{\eeq}{\end{equation}}
\newcommand{\beqa}{\begin{eqnarray}}
\newcommand{\eeqa}{\end{eqnarray}}

 \numberwithin{equation}{subsection}

\makeindex

\begin{document}

\title{\Large\textbf{Strategies and payoffs in quantum minority games}}

\author{\textit{Puya Sharif and Hoshang Heydari}\\
        \small\textit{Physics Department, Stockholm university 10691 Stockholm Sweden}\\
\\\small\textit{Email: ps@puyasharif.net}}

\maketitle

\begin{abstract}
Game theory is the mathematical framework for analyzing
strategic interactions in conflict and competition situations.
In recent years quantum game theory has earned the attention
of physicists, and has emerged as a branch of quantum information
theory \cite{invitation}. With the aid of entanglement and linear superposition
of strategies, quantum games are shown to yield significant advantage
over their classical counterparts. In this paper we explore
optimal and equilibrium solutions to quantum minority
games. Initial states with different level of entanglement
are investigated. Focus will be on 4 and 6 player games
with some $N$-player generalizations.
\end{abstract}

\section{Introduction}
Game theory is the systematic study of decision-making in strategic situations. Its models are widely
used in economics, political science, biology and computer science to capture the behavior of
individual participants in conflict and competition situations. The field attempts to describe how decision makers
do and should interact within a well-defined system of rules to maximize their payoff. The kind
of games we will be considering here is called minority games and arises in situations when a group of
non communicating agents has to independently choose between two different choices $\left|0\right\rangle $ and $\left|1\right\rangle$.
Payoff \$ of one unit goes to those agents that makes the minority choice. If agents are evenly
distributed between the two choices, everybody loses. In the classical case, the game has a mixed-strategy
solution, where agents chooses randomly between $\left|0\right\rangle$ and $\left|1\right\rangle $.
This yields an expected payoff $<\$>$ which is given by the number of combinations that results in some
player being in the minority group divided by the total number of possible combinations.
For a four player game there are 16 possible combinations, with two minority states for each player.
This gives an expected payoff <\$> of 1/8 to each.

Generally a game is defined as a set $\Gamma=\Gamma(N,\, \{s_{i}\},\,\{\$_{i}\})$, where $N$ denotes the number of players, $\{s_{i}\}$ the set of available strategies of player $i$, and $\{\$_{i}\}$ the payoffs of different game outcomes. For quantum games, we add the associated Hilbert space $\mathcal{H}$, generally of dim $2^{N}$, and the initial state $\rho$. In a quantum game, the choice of strategy $s_i$ translates to choosing a unitary operator $M_i$, which is applied locally on the qubit held by the player. The games will be analyzed with regard to two of the most important solution concepts in game theory is the Nash equilibrium and Pareto optimality. Nash equilibrium is defined as the combination of strategies $s_{i}$ for which no player gains by unilaterally changing their strategy. Pareto optimality occurs when no player can rise its payoff without lowering the payoff of others.

\section{Quantum minority games}
Following the scheme presented in \cite{Hayden}, in the quantum version of the minority game, each player is provided with a qubit from an entangled set. Strategy $s_{i}$ of player $i$ is played by doing a local unitary operation on the players own qubit, by applying its strategy operator $M\in$ SU(2). $M$ will be parameterized in the following way:

\begin{equation}
  M\left(\theta,\alpha,\beta\right)=\left(\begin{array}{cc}
e^{i\alpha}\cos\left(\theta/2\right) & ie^{i\beta}\sin\left(\theta/2\right)\\
ie^{-i\beta}\sin\left(\theta/2\right) & e^{-i\alpha}\cos\left(\theta/2\right)\end{array}\right),
  \end{equation}

with $\theta\in[0,\pi]$ and $\alpha,\beta\in[-\pi,\pi]$. The game starts out in an entangled initial state $\rho_{in}$.
\begin{equation}
\rho_{in}=\left|\psi\right\rangle \langle\psi|,
  \end{equation}
where$\left|\psi\right\rangle$  is usually taken to be a $N$ qubit GHZ-state, from which each player is provided with one qubit [2][3].
The final state $\rho_{\mathrm{fin}}$ of the game becomes
\begin{equation}
\rho_{\mathrm{fin}}=(\bigotimes_{i=1}^{N}M_{i})\rho_{\mathrm{in}}(\bigotimes_{i=1}^{N}M_{i}){}^{\dagger}.
  \end{equation}
To calculate the expected payoff of player $i$ we take the trace of the final state $\rho_{\mathrm{fin}}$ multiplied with the projection operator $P_{i}$ of the player. The projection operator projects the final state onto the desired states of player $i$.
\begin{equation}
P_{i}=\sum_{j=1}^{k}\left|\xi_{i}^{j}\right\rangle \left\langle \xi_{i}^{j}\right|.
  \end{equation}
The sum is over all the $k$ different states $\left|\xi_{i}^{j}\right\rangle$, for which player $i$ is in the minority. For $N=4$, we have the following projection operator $P_{1}$ for player 1. $P_{1}=\left|1000\right\rangle \left\langle 1000\right|+\left|0111\right\rangle \left\langle 0111\right|$. In the 6-player game, each player has a sum of 12 such states. The expected payoff $<\$>$ is finally given by:
\begin{equation}
\mathrm{<\$_{i}>=Tr[}\rho_{\mathrm{fin}}P_{i}].
  \end{equation}
The local unitary operations of the players eliminates the possibility for the
system to end up in most states where nobody wins, and therefore yields higher than classical payoff.

\subsection{Solutions with different initial states}

As a generalization of the broadly used GHZ-state as the initial state $\left|\psi_{in}\right\rangle$ we consider a superposition with products of symmetric bell pairs. A four qubit version of this state was used in a experimental implementation of a quantum minority game by C. Schmid and A.P. Flitney in \cite{Flitney exp}.
\begin{equation}
\left|\Psi(x)\right\rangle =\frac{x}{\sqrt{2}}\left|\mathrm{GHZ}_{N}\right\rangle +\sqrt{\frac{1-x^{2}}{2^{N/2}}}(\left|01\right\rangle +\left|10\right\rangle )^{\otimes N/2}.
  \end{equation}
The parameter $x\in[0,1]$ denotes the level of mixture. $x=1$ just gives back the GHZ-state and $x=0$ product of the Bell-pairs.
To account for loss in fidelity in the creation of the initial state, we form a density matrix $\rho_{in}$  out of $\mid\Psi_{in}\rangle$ and add noise that can be controlled by the parameter $f$. We get:
\begin{equation}
\rho_{in}=f\mid\Psi_{in}\rangle\langle\Psi_{in}\mid+\frac{1-f}{64}\mathbb{I_{\mathrm{64}}},
\end{equation}
 where $\mathbb{I_{\mathrm{64}}}$ is the $64\times 64$ identity matrix. By adjusting $f \rightarrow 0$, the initial state gets mixed with an even distribution of all basis states in $\mathcal{H}=(\mathbb{C}^{2})^{\otimes 6}$

 For the GHZ-state alone i.e $x=1$ and $f=1$ it has been shown that the Nash equilibrium solution $s_{NE}=M(\theta,\alpha,-\alpha)$ for the 4-player game is $M(\frac{\pi}{2},-\frac{\pi}{8},\frac{\pi}{8})$, and for the 6-player game, $M(\frac{\pi}{2},-\frac{\pi}{12},\frac{\pi}{12})$. For the state above, Schmid and Flitney showed that when starting with only the product of Bell-pairs i.e $x=0$, no advantage is achieved over the classical counterpart. For $x\leq\sqrt{\frac{2}{3}}$, a new set of Nash equilibria occurs, where the payoff is a function of $x$. This Bell-dominated region has a new Pareto optimal strategy: $M(\frac{\pi}{4},0,0)$ compared to $M(\frac{\pi}{2},-\frac{\pi}{8},\frac{\pi}{8})$ in the GHZ-dominated region $x>\sqrt{\frac{2}{3}}$.

\begin{figure}[h]
\includegraphics[scale=0.45]{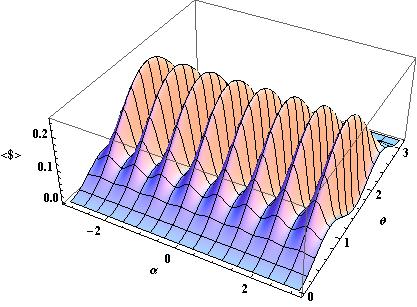}\includegraphics[scale=0.45]{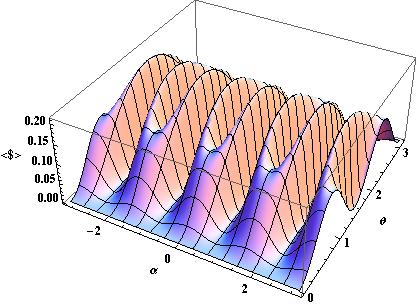}\\
\includegraphics[scale=0.45]{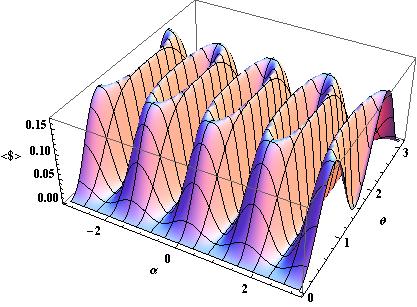}\includegraphics[scale=0.45]{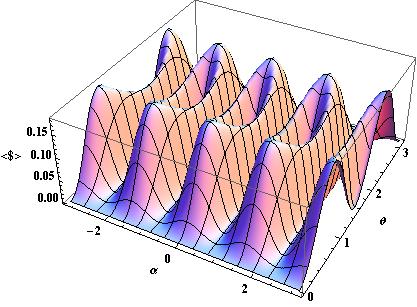}	
	\caption{Payoffs for $\hat{M}(\theta,\alpha,-\alpha).$ When $N=4$. Top left: $x=1$. Top right: $x=\sqrt{2/3}+0.1$, Bottom left: $x=\sqrt{2/3}$. Bottom right: $x=\sqrt{2/3}-0.1$}
\end{figure}

The case is different for  $N=6$, here the equilibrium strategy remains the same throughout any change of $x$. For $f=1$, the payoff is given by
\begin{equation}
<\$>=\frac{1}{4}+\frac{x^{2}}{16}.
  \end{equation}
For the pure GHZ-state this gives an equilibrium payoff of 5/16. When $x\rightarrow 0$ the payoff approaches 1/4, which is still better than the classical payoff of 3/16. This
shows that even the initial state containing only the products of Bell-pairs yields an advantage compared
to the classical expected payoff. This is not the case for general $N$. When noise is taken into account the payoff function becomes
\begin{equation}
<\$>=\frac{1}{16}(3+f+fx^{2}).
  \end{equation}
When the noise reaches maximum: $f\rightarrow 0$, the classical payoff of 3/16 returns.
It can be demonstrated that $M_{NE}=M(\frac{\pi}{2},-\frac{\pi}{12},\frac{\pi}{12})$ is a Nash equilibrium solution for all $x\in[0,1]$, by letting one player deviate from the NE solution, by playing $M_{D}(\theta^{*},-\alpha^{*},\alpha^{*})$. The following inequality holds for a Nash equilibrium:
\begin{equation}
\$_{i}(M_{NE}^{\otimes6})\geq\$_{i}(M_{D}\otimes M_{NE}^{\otimes5}).
\end{equation}

\begin{figure}[ht]
	\centering
		\includegraphics[scale=0.55]{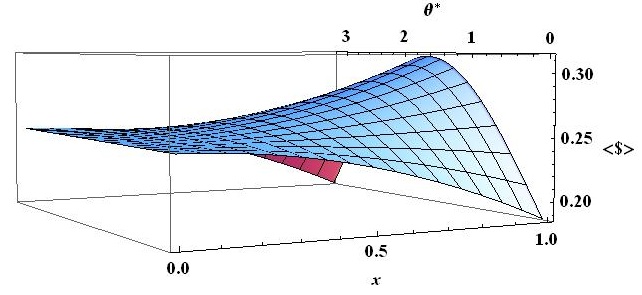}
		\caption{Payoff of a player that plays $M_{D}(\theta^{*},-\frac{\pi}{12},\frac{\pi}{12})$,
when the rest plays $M_{NE}$}.
\end{figure}

\begin{figure}[ht]
	\centering
		\includegraphics[scale=0.70]{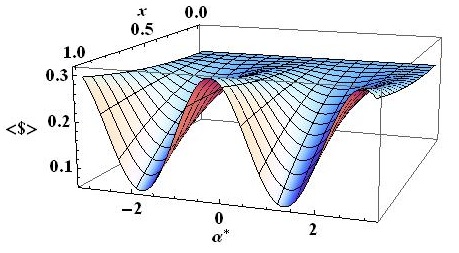}
		\caption{Payoff of a player that plays $M_{D}(\frac{\pi}{2},-\alpha^{*},\alpha^{*})$,
when the rest plays $M_{NE}$}.
\end{figure}

\subsubsection{exponential entangler}

A GHZ-state can be created by acting with an entanglement operator $J(\gamma)$ on a product state $\left|0\right\rangle \otimes\left|0\right\rangle \otimes\left|0\right\rangle \otimes\cdots\otimes\left|0\right\rangle$ , where
\begin{equation}
J(\gamma)=\mathrm{exp\left(i\frac{\gamma}{2}\sigma_{x}^{\otimes\mathit{N}}\right)}.
  \end{equation}
We then have
\begin{equation}
\left|\Psi(\gamma)\right\rangle =J\left|00\cdots0\right\rangle ,
  \end{equation}
where $\gamma\in[0,\frac{\pi}{2}]$ is a parameter that controls the level of entanglement. This gives an output state of the following form
\begin{equation}
\left|\Psi(\gamma)\right\rangle =\cos(\frac{\gamma}{2})\left|00\cdots0\right\rangle +i\sin(\frac{\gamma}{2})\left|11\cdots1\right\rangle .
  \end{equation}
Maximum is reached for $\gamma=\frac{\pi}{2}.$ If $\left|\Psi(\gamma)\right\rangle$  is used as initial state for a quantum minority game, the Nash equilibrium payoffs will depend on the parameter $\gamma$.
For $\gamma=0$ the classical payoffs are obtained, since the game starts out in an unentangled initial state \cite{decoh}. A $N$-player generalization has been conjectured:
\begin{equation}
<\$>_{N}=\left(<\$>_{C}-\frac{1}{2}<\$>_{Q}\right)\left(\cos\frac{\gamma}{2}-\sin\frac{\gamma}{2}\right)^{2}+\frac{1}{2}<\$>_{Q}\left(\cos\frac{\gamma}{2}+\sin\frac{\gamma}{2}\right)^{2},
  \end{equation}
where $<\$>_{C}$ is the classically obtainable payoffs for for classical NE strategies, and $<\$>_{C}$ for the quantum versions \cite{Chen}.

\subsubsection{Products of W-states}
A six player game could use a product of two three qubit W-states as its initial state $\left|\psi_{in}\right\rangle$.
\begin{equation}
\left|\psi_{in}\right\rangle =\left|W_{3}\right\rangle \otimes\left|W_{3}\right\rangle ,
  \end{equation}
where
\begin{equation}
\left|W_{3}\right\rangle =\frac{1}{\sqrt{3}}(\left|001\right\rangle +\left|010\right\rangle +\left|100\right\rangle ).
  \end{equation}
$\left|\psi_{in}\right\rangle$  is a symmetric superposition of nine states with four qubits in the $\left|0\right\rangle$ -state and two in the $\left|1\right\rangle$ -state, compactly written as $\left|\psi_{in}\right\rangle =\left|4,2\right\rangle$. This state therefore has tree minority combinations for each player, and no undesired states! The game simply starts out in the best possible configuration, and the only thing the players should do is to apply the identity operator $I$, to obtain an expected payoff 1/3, the theoretical maximum for a six-player game. This solution Pareto optimal, compared to the six-player game starting with an GHZ-state, which is not.

\end{document}